\documentclass[pdflatex,sn-mathphys-num]{sn-jnl}

\usepackage{graphicx}
\usepackage{multirow}
\usepackage{amsmath,amssymb,amsfonts}
\usepackage{amsthm}
\usepackage{mathrsfs}
\usepackage[title]{appendix}
\usepackage{xcolor}
\usepackage{textcomp}
\usepackage{manyfoot}
\usepackage{booktabs}
\usepackage{algorithm}
\usepackage{algorithmicx}
\usepackage{algpseudocode}
\usepackage{listings}

\theoremstyle{thmstyleone}

\theoremstyle{thmstyletwo}

\theoremstyle{thmstylethree}

\begin{document}

\title[Article Title]{Recent progress on inflation and dark energy from string theory}

\author[1,2]{\fnm{Michele} \sur{Cicoli}}

\affil[1]{\orgdiv{Dipartimento di Fisica e Astronomia}, \orgname{Università di Bologna}, \orgaddress{\street{via Irnerio 46}, \city{Bologna}, \postcode{40126}, \country{Italy}}}

\affil[2]{\orgdiv{INFN}, \orgname{Sezione di Bologna}, \orgaddress{\street{viale Berti Pichat 6/2}, \city{Bologna}, \postcode{40127}, \country{Italy}}}

\abstract{We review recent progress in string model building in both early and late time cosmology. We describe the main theoretical and phenomenological features of an entire class of inflationary models where inflation is driven by a type IIB K\"ahler modulus which enjoys an effective and approximate shift symmetry. We illustrate how reheating can occur via the perturbative decay of the modulus into visible and hidden sector degrees of freedom, paying particular attention to the associated production of axionic dark radiation. We quickly discuss the status of de Sitter vacua versus quintessence model building in string theory, analysing the level of control of these constructions and the main challenges faced by models of dynamical dark energy. We finally present a working model of axion hilltop quintessence in string theory, stressing the importance of initial conditions.}

\keywords{String inflation, String dark energy}

\maketitle

\section{Introduction}
\label{sec1}

Cosmology is a natural playground for testing string theory. This is due to the fact that understanding crucial epochs of the cosmological evolution of our Universe necessarily requires to control Planck-scale physics. Some examples are inflation, which is notoriously sensitive to $M_p$-suppressed operators, eras when scalar fields roll through field ranges which are multiples of $M_p$, or perturbative reheating which is dominated by the decay of the longest lived particles that are naturally those which couple only gravitationally to ordinary matter. 

Considerable progress has been made on different aspect of string cosmology (for a comprehensive review see \cite{Cicoli:2023opf}) and many models are present in the literature. In this note we focus on a particularly promising class of models within the framework of type IIB flux compactifications on Calabi-Yau orientifolds where moduli stabilisation is best understood. In this context the main characters of the history of our Universe are the K\"ahler moduli $T_i=\tau_i + i \theta_i$. 

We will argue that the saxions $\tau_i$, which are the scalar partners of the axions $\theta_i$, are good inflaton candidates since they enjoy an approximate rescaling shift symmetry that is broken only by quantum effects. Generically the perturbative decay of these saxions drive reheating with the potential production of ultra-light axions $\theta_i$ that contribute to dark radiation.

Models of dynamical dark energy seem harder to build than de Sitter vacua. However, if quintessence will turn out to be preferred by observations, the main candidates to drive a present epoch of accelerated expansion seem to be the axions $\theta_i$ since they enjoy a shift symmetry that is exact at perturbative level. After a general discussion, we will sketch the main features of a working string model that can describe the history of our Universe from inflaton to quintessence \cite{Cicoli:2024yqh}.

\section{Inflation}
\label{Infl}

\subsection{Single-field slow-roll}

\begin{figure}[h]
\centering
\includegraphics[width=0.75\textwidth]{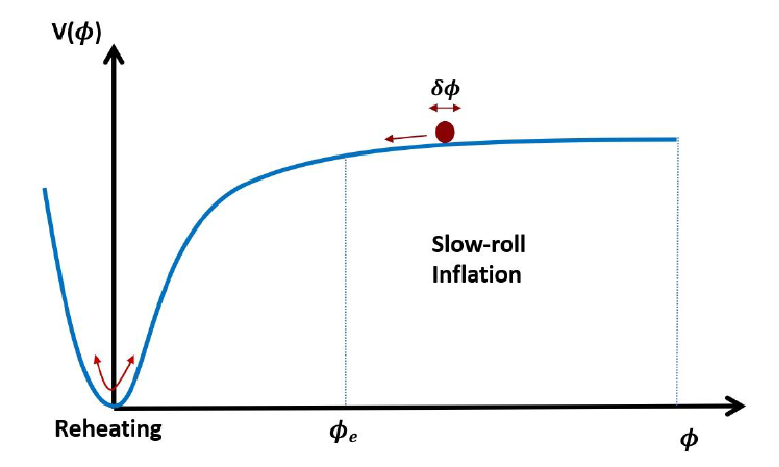}
\caption{A typical single-field slow-roll potential taken from \cite{Cicoli:2023opf}.}
\label{Fig1}
\end{figure}

A plethora of models of inflation is present in the literature with several examples involving a complicated multifield dynamics. However, as far as matching data is concerned, there is no need to invoke sophisticated constructions since simple single-field slow-roll inflation seems to work rather well. The typical potential of these models is characterised by a flat plateau, and can be written as (see Fig. \ref{Fig1}):
\begin{equation}
V(\phi) = V_0 \left[1- g(\phi)\right] \simeq V_0\quad\text{since}\quad g(\phi)\ll1\quad\text{for}\quad \phi\gg 1\,,
\label{Vslow-roll}
\end{equation}
where everything is measured in units of the reduced Planck mass $M_p=1/\sqrt{8\pi G_N}\simeq 2.4\times 10^{18}$ GeV.

\subsection{Inflating with string moduli}

String compactifications typically lead to hundreds of fields in the low-energy 4D effective field theory. However most of them, like the complex structure moduli and the axio-dilaton of type IIB constructions, become very heavy due to flux stabilisation. The remaining K\"ahler moduli can then lead to an inflationary dynamics that is effectively single-field and reproduces the successful picture of slow-roll inflation with a potential given by (\ref{Vslow-roll}). 

A ubiquitous K\"ahler modulus is the Calabi-Yau volume mode $\mathcal{V}$ which couples to all sources of energy, as can be easily seen from the $e^K=\mathcal{V}^{-2}$ factor in front of the supergravity F-term scalar potential. Hence one cannot have an inflaton-independent plateau if the inflaton coincides with $\mathcal{V}$. This implies that the inflaton should be a direction in K\"ahler moduli space orthogonal to $\mathcal{V}$. In what follows we shall call this canonically unnormalised field $\tau_\phi$, leaving $\phi$ to denote the canonically normalised inflaton.

Since each term in $V$ depends on $\mathcal{V}$, $V(\phi)\simeq V_0$ only if the leading dynamics fixes $\mathcal{V}$ but not $\tau_\phi$. This is possible only if $\tau_\phi$ is a leading order flat direction with an approximate shift symmetry. As argued in \cite{Burgess:2014tja,Burgess:2016owb}, this picture is indeed realised in the type IIB K\"ahler sector as follows. At tree-level, the famous no-scale cancellation makes the potential for \textit{all} K\"ahler moduli flat. As confirmed by several studies, the leading no-scale breaking effects are $\mathcal{O}(\alpha'^3)$ corrections which lift only $\mathcal{V}$ \cite{Becker:2002nn}. All the other K\"ahler moduli appear in general in $V$ via string loop corrections which however enjoy a 1-loop extended no-scale \cite{Cicoli:2007xp}, making all K\"ahler moduli orthogonal to $\mathcal{V}$ good candidates to realise slow-roll inflaton along a plateau. Clearly, a crucial ingredient that has to be controlled to achieve a working model is the size of subdominant quantum effects that determine the length of the plateau which, in turn, affects the duration of inflation.

\subsubsection*{Inflationary dynamics}

Schematically, if we focus for simplicity on a two-field model with the volume $\mathcal{V}$ and the inflaton $\tau_\phi$, the total potential looks like: 
\begin{equation}
V_{\rm tot}(\mathcal{V},\tau_\phi) = V_{\rm lead}(\mathcal{V})-V_{\rm sub}(\mathcal{V},\tau_\phi)\,,
\end{equation}
where $V_{\rm lead}(\mathcal{V})$ is the leading order potential which generates a minimum for the volume direction, while $V_{\rm sub}(\mathcal{V},\tau_\phi)\ll V_{\rm lead}(\mathcal{V})$ is the subdominant contribution which lifts the inflaton direction and produces just a small inflaton-dependent shift of the minimum for $\mathcal{V}$. Ignoring this small shift, the volume can be set at its minimum, $\mathcal{V}=\langle\mathcal{V}\rangle$, and the total potential for the canonically normalised inflaton $\phi$ takes exactly the same form as in (\ref{Vslow-roll}) with:
\begin{equation}
V_0\equiv V_{\rm sub}(\langle\mathcal{V}\rangle, \langle\tau_\phi\rangle),
\end{equation}
to guarantee an overall Minkowski minimum and:
\begin{equation}
g(\phi)\equiv \frac{V_{\rm sub}(\langle\mathcal{V}\rangle, \tau_\phi(\phi))}{V_{\rm sub}(\langle\mathcal{V}\rangle, \langle\tau_\phi\rangle)}\,,
\end{equation}
with $\tau_\phi(\phi)$ determined by canonical normalisation. Since $\tau_\phi$ is a leading order flat direction, $V_{\rm sub}$ should tend to zero for large field values, implying:
\begin{equation}
V_{\rm sub}(\langle\mathcal{V}\rangle, \tau_\phi)\ll V_{\rm sub}(\langle\mathcal{V}\rangle, \langle\tau_\phi\rangle)\qquad\text{for}\qquad\tau_\phi>\langle\tau_\phi\rangle\,,
\end{equation}
which, in turn, implies that $g(\phi)\ll 1$ and $V(\phi)\simeq V_0$ for $\phi\gg 1$.

\subsection{String inflation potentials}

It is important to stress that the function $g(\phi)$, which determines the shape of the inflationary potential (\ref{Vslow-roll}), depends on two features:
\begin{enumerate}
\item The origin of the effects generating $V_{\rm sub}(\langle\mathcal{V}\rangle, \tau_\phi)$ that can be of two types:
\begin{itemize}
\item \emph{Perturbative effects:} Corrections which are perturbative in both $g_s$ and $\alpha'$ are typically power-law and vanish at large field values:
\begin{equation}
V_{\rm sub}(\langle\mathcal{V}\rangle, \tau_\phi)   \propto \frac{1}{\tau_\phi^p}\to 0\qquad\text{for}\qquad \tau_\phi\to \infty\qquad\text{and}\qquad p>0\,.
\end{equation}
\item \emph{Non-perturbative effects:} At non-perturbative level the effective action receives corrections which are typically exponential in the K\"ahler moduli and scale as:
\begin{equation}
V_{\rm sub}(\langle\mathcal{V}\rangle, \tau_\phi)   \propto e^{-k\tau_\phi}\to 0\qquad\text{for}\qquad \tau_\phi\to \infty\qquad\text{and}\quad k>0\,.
\end{equation}
\end{itemize}

\item The relation $\tau_\phi(\phi)$ that expresses the K\"ahler modulus in terms of the canonically normalised inflaton $\phi$ which depends on the topology of the divisor whose volume is controlled by $\tau_\phi$. One can have in general two cases:
\begin{itemize}
\item \emph{Bulk (fibre) modulus:} When the inflaton is a bulk divisor, as a standard modulus in supergravity, its canonical normalisation is exponential:
\begin{equation}
\tau_\phi = e^{\lambda\phi}\qquad\text{with}\qquad\lambda\sim\mathcal{O}(1)\,.
\label{CanNormBulk}
\end{equation}
\item \emph{Local (blow-up) modulus:} When the inflaton is topologically the resolution of a point-like CY singularity, the canonical normalisation becomes instead power-law: 
\begin{equation}
\tau_\phi =  \mu\,\mathcal{V}^{2/3}\phi^{4/2}   \qquad\text{with}\qquad\mu\sim\mathcal{O}(1)\,.
\end{equation}
\end{itemize}
\end{enumerate}

Four different classes of models have been constructed in the literature by combining these two features:
\begin{itemize}
\item \emph{Non-perturbative blow-up inflation:} Historically, the first class of models involved a blow-up inflaton with a potential generated by non-perturbative effects \cite{Conlon:2005jm,Bond:2006nc}. In this case $g(\phi)$ takes the form:
\begin{equation}
g(\phi)\sim e^{-k\mu\,\mathcal{V}^{2/3}\phi^{4/3}}\ll1 \qquad\text{for}\qquad\phi>0\,.
\end{equation}
\item \emph{Non-perturbative fibre inflation:} Non-perturbative effects can induce a very flat plateau also for fibre moduli \cite{Cicoli:2011ct,Lust:2013kt} with:
\begin{equation}
g(\phi)\sim e^{-k e^{\lambda\phi}}\ll1 \qquad\text{for}\qquad\phi>0\,.
\end{equation}
\item \emph{Loop fibre inflation:} A promising class of models relies on potentials generated by loop corrections for fibre moduli \cite{Cicoli:2008gp,Broy:2015zba,Cicoli:2016chb}. After canonical normalisation, $g(\phi)$ becomes exponential also in this case:
\begin{equation}
g(\phi)\sim e^{-p\lambda\phi}\ll1 \qquad\text{for}\qquad\phi>0\,.
\end{equation}
Note that this is an $\alpha$-attractor realisation in string theory \cite{Kallosh:2017wku}.
\item \emph{Loop blow-up inflation:} Interestingly, it has been realised that loop corrections for blow-up modes can yield a power-law inflationary potential of the form \cite{Bansal:2024uzr}:
\begin{equation}
g(\phi)\sim \frac{1}{\mathcal{V}^{2p/3}\phi^{4p/3}}\ll1 \qquad\text{for}\qquad\phi\lesssim 1\,.
\end{equation}
The leading known loop corrections to the K\"ahler potential from string theory give $p=1/2$ that translates into a potential of the form:
\begin{equation}
V=V_0\left(1-\frac{c}{\mathcal{V}^{1/3}\phi^{2/3}}\right).
\end{equation}
\end{itemize}
In what follows, we shall briefly summarise the main features of this model as an illustrative example.

\subsection{The loop blow-up inflation model}

Recall that the K\"ahler moduli are complex fields $T_i=\tau_i+i\theta_i$. Let us focus on a simple type IIB compactification on a CY threefold with volume:
\begin{equation}
\mathcal{V}=\tau_b^{3/2}-\tau_s^{3/2}-\tau_\phi^{3/2}\simeq \tau_b^{3/2}\,.    
\end{equation}
The low-energy effective field theory is characterised by a K\"ahler potential (at tree- and $\alpha'^3$-level) and a superpotential (at tree- and non-perturbative-level) of the form:
\begin{equation}
K =-2\ln\left(\mathcal{V}+\frac{\xi}{g_s^{3/2}}\right)
\qquad W=W_0+ A_s\,e^{-a_s T_s}+ A_\phi\, e^{-a_\phi T_\phi}\,,
\end{equation}
where $\xi$, $W_0$, $A_s$, $A_\phi$ and $a_\phi>a_s$ are $\mathcal{O}(1)$ positive constants, and the string coupling $g_s$ is in the perturbative regime, i.e. $g_s\ll 1$. Using the standard expression for the supergravity F-term potential, the scalar potential arising from $K$ and $W$ features a leading and a subleading term: 
\begin{equation}
V=V_{\rm lead}(\mathcal{V},\tau_s)+V_{\rm sub}(\mathcal{V},\tau_\phi)\,,
\end{equation}
where:   
\begin{eqnarray}
V_{\rm lead}(\mathcal{V},\tau_s)&=& \frac{C_{\rm up}}{\mathcal{V}^2}+C_s\,\frac{\sqrt{\tau_s}\,e^{-2 a_s\tau_s}}{\mathcal{V}} -D_s\,\frac{\tau_s\,e^{-a_s\tau_s}}{\mathcal{V}^2}+\frac{C_{\alpha'}}{g_s^{3/2}\mathcal{V}^3}\,,
\label{Vlead}\\
V_{\rm sub}(\mathcal{V},\tau_\phi)&=& C_\phi\,\frac{\sqrt{\tau_\phi}\,e^{-2 a_\phi\tau_\phi}}{\mathcal{V}} -D_\phi\,\frac{\tau_\phi\,e^{-a_\phi\tau_\phi}}{\mathcal{V}^2}\,.
\label{Vsub}
\end{eqnarray}
In (\ref{Vlead}) we have introduced a generic uplifting contribution proportional to $C_{\rm up}$ which is needed to obtain a Minkowski vacuum. Moreover, $C_s$, $D_s$, $C_{\alpha'}$, $C_\phi$ and $D_\phi$ are all $\mathcal{O}(1)$ positive constants. Note that, in the region of our interest where $\tau_\phi\gtrsim \tau_s$, (\ref{Vsub}) is exponentially suppressed with respect to (\ref{Vlead}) since $a_\phi>a_s$. This guarantees that, at leading order, the minimum for $\mathcal{V}$ and $\tau_s$ is not affected by the evolution of $\tau_\phi$ during the inflationary dynamics. After the end of inflation all fields relax to the global Minkowski minimum that is located schematically at:
\begin{equation}
\langle\tau_s\rangle\sim g_s^{-1}\qquad\text{and}\qquad \langle\mathcal{V}\rangle\sim e^{a_s\langle\tau_s\rangle}\sim e^{a_\phi\langle\tau_\phi\rangle}\,.
\end{equation}

\subsection{Loop corrections}

Inflation occurs along the flat plateau in the region $\tau_\phi >\langle\tau_\phi\rangle$ where the potential (\ref{Vsub}) becomes exponentially suppressed. In this region loop corrections can be important since they tend to zero more slowly than non-perturbative effects due to the fact that they are power-law in $\tau_\phi$. It is therefore crucial to understand the behaviour of loop corrections to the K\"ahler potential. At 1-loop level, these perturbative corrections have been computed explicitly just for toroidal orientifolds \cite{Berg:2005ja}. In this case the volume is $\mathcal{V}=\sqrt{\tau_1\tau_2\tau_3}$ and 1-loop corrections to $K$ receive two contributions:
\begin{equation}
\qquad \delta K_{(g_s)}= \delta K_{(g_s)}^{KK} + \delta K_{(g_s)}^W\,.
\end{equation}
The first term arises from the tree-level exchange of Kaluza-Klein (KK) closed strings between parallel D7-branes/O7-planes and reads:
\begin{equation}
\delta K_{(g_s)}^{KK}= g_s\left(\frac{C_1^{KK}(U,\overline{U})}{\tau_1}+\frac{C_2^{KK}(U,\overline{U})}{\tau_2}+\frac{C_3^{KK}(U,\overline{U})}{\tau_3}\right).  
\end{equation}
The second contribution is due to the tree-level exchange of winding (W) closed strings at the intersection between different stacks of D7-branes and looks like:
\begin{equation}
\delta K_{(g_s)}^{W}= \frac{C_1^{W}(U,\overline{U})}{\tau_2\tau_3}+\frac{C_2^{W}(U,\overline{U})}{\tau_1\tau_3}+\frac{C_3^{W}(U,\overline{U})}{\tau_1\tau_2}\,.
\end{equation}
These results have been generalised to an arbitrary CY compactification with the following conjecture \cite{Berg:2007wt}:
\begin{eqnarray}
\delta K_{(g_s)}^{KK}&=& \frac{g_s}{\mathcal{V}}\sum_i C_i^{KK}(U,\overline{U})\,M^{-2}_{KK,i}\,,\\
\delta K_{(g_s)}^{W}&=&   \frac{1}{\mathcal{V}}\sum_i C_i^W(U,\overline{U})\, M^{-2}_{i,W}\,,
\end{eqnarray}
where $C_i^{KK}$ and $C_i^W$ are expected to be $\mathcal{O}(1)$ functions of the complex structure moduli $U$, and $M_{KK,i}$ and $M_{W,i}$ are the masses, respectively, of the exchanged KK and winding modes. Interestingly, once $M_{KK,i}$ and $M_{W,i}$ are written in terms of the K\"ahler moduli, one can show in full generality that these 1-loop corrections to $K$ enjoy an extended no-scale cancellation in the scalar potential for an arbitrary CY compactification \cite{Cicoli:2007xp}. Hence they turn out to be subdominant with respect to the non-perturbative potential (\ref{Vlead}) around the minimum. As we shall see, this is however not true in the inflationary region. 

\subsection{Loop corrections from effective field theory}

The conjectured behaviour of loop corrections can be strengthened using effective field theory arguments \cite{vonGersdorff:2005bf,Cicoli:2007xp,Gao:2022uop,Bansal:2024uzr}. In supersymmetric theories, 1-loop corrections to the K\"ahler potential induce corrections to both the kinetic terms and the scalar potential. 
Focusing on 1-loop corrections to the 2-point function of a light mode $L$ coupled to a heavy mode $H$ as:
\begin{equation}
\mathcal{L}\supset M^2 H^2 + g L H^2\,,  
\end{equation}
the 2-point function 1-loop renormalisation would induce a 1-loop correction to the kinetic terms of $L$ of the form:
\begin{equation}
\mathcal{L}_{\rm kin} \simeq\left[1+c_{\rm loop}\left(\frac{g}{M}\right)^2\right]\partial_\mu L\partial^\mu L\,.   
\label{1loopL}
\end{equation}
where $c_{\rm loop}$ is a loop-factor that is expected to be of order $c_{\rm loop}\simeq 1/(16\pi^2)$. When $H$ is a winding, Kaluza-Klein or massive string mode, and $L$ is a K\"ahler modulus $\tau$ controlling the overall volume, i.e. $\tau=\mathcal{V}^{2/3}$, the coupling $g$ arises from the $\tau$-dependence of the mass of $H$ as follows:
\begin{equation}
M^2(\tau) H^2 = M^2(\langle\tau\rangle) H^2 + \frac{\partial M^2}{\partial \tau}(\langle\tau\rangle) \left(\tau-\langle\tau\rangle\right) H^2 + \cdots 
\end{equation}
As we have seen in (\ref{CanNormBulk}), the canonical normalisation of a bulk modulus is exponential, implying:
\begin{equation}
\frac{\left(\tau - \langle\tau\rangle\right)}{\langle\tau\rangle} \simeq \frac{L}{M_p} \,.
\label{UsefulCanNorm}
\end{equation}
Given that the mass of all relevant heavy modes is power-law in $\tau$, one ends up with:
\begin{equation}
\frac{\partial M^2}{\partial \tau}(\langle\tau\rangle) \left(\tau - \langle\tau\rangle\right)\simeq M^2(\langle\tau\rangle) \frac{\left(\tau - \langle\tau\rangle\right)}{\langle\tau\rangle}\qquad\Rightarrow \qquad g\simeq \frac{M^2}{M_p}\,.  
\end{equation}
Substituting this result in (\ref{1loopL}), we can compare the kinetic Lagrangian with the one obtained in terms of the corrected K\"ahler potential $K=K_0+\delta K$:
\begin{equation}  
\mathcal{L}_{\rm kin} \simeq \left[\frac{\partial^2 K_0}{\partial\tau^2}+ \frac{\partial^2 \delta K}{\partial\tau^2}\right]\partial_\mu \tau \partial^\mu \tau \simeq 
\left[1 + \frac{\partial^2 (\delta K)}{\partial\tau^2} \tau^2\right]\partial_\mu L\partial^\mu L\,,
\end{equation}
where we used (\ref{UsefulCanNorm}). Hence we infer:
\begin{equation}
\frac{\partial^2 (\delta K)}{\partial\tau^2}\simeq \frac{c_{\rm loop}}{\tau^2}\left(\frac{g}{M}\right)^2 \qquad\Rightarrow\qquad \delta K \simeq c_{\rm loop}\left(\frac{M}{M_p}\right)^2\,.
\end{equation}
Different 1-loop contributions can therefore arise depending on the nature of the heavy modes running in the loop:
\begin{itemize}
\item If $H$ is a massive string state:
\begin{equation}
M\equiv M_s\simeq \frac{M_p}{\sqrt{\mathcal{V}}} \quad\Rightarrow\quad\delta K\simeq \frac{c_{\rm loop}}{\mathcal{V}}\,,
\end{equation}
which matches the $\mathcal{V}$-scaling of the $\mathcal{O}(\alpha'^3)$ term derived in \cite{Becker:2002nn}.

\item If $H$ is a winding mode:
\begin{equation}
M\equiv M_{W}\simeq \frac{M_p}{\sqrt{\mathcal{V}}}\,\tau^{1/4} \quad\Rightarrow\quad\delta K\simeq K\simeq c_{\rm loop}\frac{\sqrt{\tau}}{\mathcal{V}}\,,
\end{equation}
which matches the K\"ahler moduli dependence of the correction $\delta K_{(g_s)}^{KK}$ conjectured in \cite{Berg:2007wt}. Hence $\delta K_{(g_s)}^{KK}$ can be seen to arise from the tree-level exchange of KK closed string or from 1-loops of winding open strings. These corrections would however feature an extended no-scale cancellation \cite{Cicoli:2007xp}.

\item If $H$ is a Kaluza-Klein mode:
\begin{equation}
M\equiv M_{KK}\simeq \frac{M_p}{\sqrt{\mathcal{V}}\,\tau^{1/4}} \quad\Rightarrow\quad\delta K\simeq \frac{c_{\rm loop}}{\mathcal{V}\sqrt{\tau}}\,,
\end{equation}
which matches the K\"ahler moduli dependence of $\delta K_{(g_s)}^W$ found in \cite{Berg:2007wt}. Hence $\delta K_{(g_s)}^W$ can be seen to arise from the tree-level exchange of winding closed string or from 1-loops of KK open strings. Note that, if $\tau =\tau_\phi$, this correction to $K$ would induce a correction to the scalar potential of the form:
\begin{equation}
\delta K\simeq \frac{c_{\rm loop}}{\mathcal{V}\sqrt{\tau_\phi}} \quad\Rightarrow\quad \delta V\simeq \frac{c_{\rm loop}}{\mathcal{V}^3\sqrt{\tau_\phi}}\,.
\label{V1loop}
\end{equation}
This would become the leading inflaton-dependent correction to $V$ which is crucial for inflation.
\end{itemize}

Another effective field theory argument in favour of the form of the 1-loop correction (\ref{V1loop}) comes from the observation that this correction should reproduce the well-known 1-loop Coleman-Weinberg potential \cite{Cicoli:2007xp}:
\begin{equation}
V^{\rm CW}_{\rm 1-loop} \simeq \frac{1}{16\pi^2}\Lambda^2\text{Str}\,\mathcal{M}^2\,,
\end{equation}         
where in supergravity the supertrace scales in terms of the gravitino mass $m_{3/2}$ as:
\begin{equation}
\text{Str}\,\mathcal{M}^2\simeq m_{3/2}^2\simeq\frac{M_p^2}{\mathcal{V}^2}\,.
\end{equation} 
The cut-off $\Lambda$ is expected instead to be given by the mass of Kaluza-Klein replicas of open strings on D7-branes. There are two cases:
\begin{enumerate}
\item D7-branes on $\tau_b$:
\begin{equation}
\Lambda\simeq\frac{M_p}{\mathcal{V}^{2/3}}\quad\Rightarrow\quad\delta V_{(g_s)}\simeq\frac{c_{\rm loop}}{\mathcal{V}^{10/3}}\,,
\end{equation}
which is subdominant with respect to (\ref{V1loop}).

\item D7-branes on $\tau_s$:
\begin{equation}
\Lambda\simeq\frac{M_p}{\tau_\phi^{1/4}\sqrt{\mathcal{V}}}\quad\Rightarrow\quad\delta V_{(g_s)}\simeq\frac{c_{\rm loop}}{\mathcal{V}^3\sqrt{\tau_\phi}}\,,
\end{equation}
\end{enumerate}
which is the same expression as (\ref{V1loop}) that has been inferred from the 1-loop renormalisation of the 2-point function of the volume mode. Note that, if there are no D7-branes wrapped on $\tau_\phi$, one can still have KK modes of $\tau_\phi$ (closed strings) running in loops, giving again:
\begin{equation}
\Lambda\simeq\frac{M_p}{\tau_\phi^{1/4}\sqrt{\mathcal{V}}}\,.
\end{equation}
This argument implies that $\tau_\phi$-dependent loop corrections to $V$ are unavoidable.

\subsection{Inflaton potential}

\begin{figure}[h]
\centering
\includegraphics[width=0.85\textwidth]{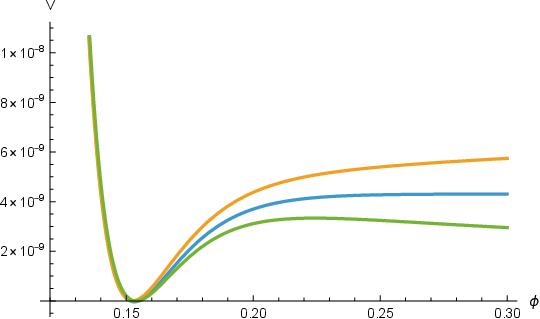}
\caption{Inflationary potential for the canonically normalised inflaton $\phi$ with $C_\phi=D_\phi=a_\phi=1$ and $\beta$ chosen appropriately to obtain a Minkowski vacuum. The blue line corresponds to $c_{\rm loop}=0$, the orange one to $c_{\rm loop}=10$ and the green one to $c_{\rm loop}=-10$.}
\label{Fig2}
\end{figure}

Let us now include the leading order loop correction to the scalar potential of the loop blow-up inflation model. The total inflationary potential including string loops takes the form:
\begin{equation}
V(\tau_\phi) = \frac{\beta}{\mathcal{V}^3}+C_\phi\frac{\sqrt{\tau_\phi}\,e^{-2 a_\phi\tau_\phi}}{\mathcal{V}}-D_\phi\,\frac{\tau_\phi\,e^{-a_\phi\tau_\phi}}{\mathcal{V}^2}-\frac{c_{\rm loop}}{\mathcal{V}^3\sqrt{\tau_\phi}}\,,
\end{equation}
where $\tau_\phi$ should be written in terms of the canonically normalised inflaton as:
\begin{equation}
\phi=\sqrt{\frac{4}{3\mathcal{V}}}\tau_\phi^{3/4}\,.
\end{equation}
An illustrative plot of the inflationary potential is presented in Fig. \ref{Fig2}

As estimated in \cite{Bansal:2024uzr}, in order to reproduce the dynamics of Non-perturbative blow-up inflation \cite{Conlon:2005jm} where the loop correction is negligible throughout the whole duration of inflation, the coefficient $c_{\rm loop}$ should be tiny: $c_{\rm loop}\lesssim 10^{-6}$. Given that this coefficient is expected to be much larger, of order $c_{\rm loop}\simeq 1/(16\pi^2)$, a more realistic scenario is the one where in the plateau the loop correction dominates over non-perturbative effects, leading to a potential that in the inflationary region can be approximated as:
\begin{equation}
V(\tau_\phi) \simeq \frac{\beta}{\mathcal{V}^3}-\frac{c_{\rm loop}}{\mathcal{V}^3\sqrt{\tau_\phi}} = V_0\left(1-\frac{c_{\rm loop}}{\mathcal{V}^{1/3}\phi^{2/3}}\right)\quad\text{with}\quad V_0\equiv\frac{\beta}{\mathcal{V}^3}\,.
\label{VinfApp}
\end{equation}
Note that this potential would yield a viable slow-roll dynamics towards the minimum only for $c_{\rm loop}>0$.

\subsection{Inflationary dynamics}

Assuming $c_{\rm loop}>0$, the slow-roll parameters associated to the Loop blow-up inflation potential (\ref{VinfApp}) look like:
\begin{eqnarray}
\epsilon&=& \frac12 \left(\frac{V_\phi}{V}\right)^2 \simeq\frac29 \frac{c_{\rm loop}^2}{\mathcal{V}^{2/3}\phi^{10/3}}\,,  \\
\eta&=& \frac{V_{\phi\phi}}{V}\simeq -\frac{10}{9}\frac{c_{\rm loop}}{\mathcal{V}^{1/3}\phi^{8/3}}\,.
\end{eqnarray}
Consequently, the main cosmological observables read:
\begin{eqnarray}
N_e&=&\int_{\phi_{\rm end}}^{\phi_*}\frac{V}{V_\phi}d\phi\simeq \frac{9}{16}\frac{\mathcal{V}^{1/3}\phi_*^{8/3}}{c_{\rm loop}}\,, \\
\hat{A}_s&=&\frac{9 V_0}{4}\frac{\mathcal{V}^{2/3}\phi_*^{10/3}}{c_{\rm loop}^2}\simeq 2.7\times 10^{-7}\,.
\end{eqnarray}
Setting $c_{\rm loop}=1/(16\pi^2)$, the previous relations can be used to obtain:
\begin{equation}
\phi_*\simeq 0.06\,N_e^{7/22}\qquad\text{and}\qquad \mathcal{V}\simeq 1743\,N_e^{5/11}\,.
\end{equation}
These can be used to obtain the expressions for the scalar spectral index $n_s$ and the tensor-to-scalar-ratio $r$ in terms of the number of e-foldings $N_e$ as:
\begin{eqnarray}
n_s&=& 1+2\eta-6\epsilon \simeq 1-\frac{20}{9}\frac{c_{\rm loop}}{\mathcal{V}^{1/3}\phi_*^{8/3}} \simeq 1-\frac{1.25}{N_e}\,, \label{ns} \\ 
r&=&16\epsilon\simeq \frac{32}{9}\frac{c_{\rm loop}^2}{\mathcal{V}^{2/3}\phi_*^{10/3}}\simeq \frac{0.004}{N_e^{15/11}}\,.
\label{r}
\end{eqnarray}
These two expressions can be combined to find an interesting relation between $n_s$ and $r$ that is typical of the Loop blow-up inflation model:
\begin{equation}
r\simeq 0.003\left(1-n_s\right)^{15/11}\,.
\end{equation}
The number of e-foldings $N_e$ is determined by the post-inflationary evolution of the model which in our case, depending on the details of the Standard Model realisation on branes, yields $51.5\lesssim N_e\lesssim 53$. This, in turn, leads to the prediction $0.9757\lesssim n_s\lesssim 0.9764$  which is in very good agreement with recent CMB and BAO data that give $n_s=0.9728\pm 0.0029$ ($68\%$ CL) \cite{Balkenhol:2025wms}. Moreover, the tensor-to-scalar ratio $r$ is predicted to be of order $r\simeq 2\times 10^{-5}$. This value is well below the present observational upper bound, $r<0.034$ ($95\%$ CL) \cite{Balkenhol:2025wms}, and is beyond the reach of polarisation observations that will be performed in the near future.

\subsection{Control over effective field theory}

An important UV requirement to check is if the whole inflationary dynamics can take place remaining in a region of moduli space where the effective field theory is under control. This depends on the number of e-foldings $N_e$. In fact, once $N_e$ is fixed, one can derive the required values of the UV parameters as:
\begin{equation}
\mathcal{V}\simeq 1743\,N_e^{5/11}\sim\mathcal{O}(10^4)\quad\text{and}\quad \phi_*\simeq 0.06\,N_e^{7/22}\sim \mathcal{O}(0.2)\quad \text{for}\quad 51.5\lesssim N_e\lesssim 53\,. \nonumber
\end{equation}
From the canonical normalisation:
\begin{equation}
\phi\simeq \frac{\tau_\phi^{3/4}}{\sqrt{\mathcal{V}}}\simeq\left(\frac{\tau_\phi}{\tau_b}\right)^{3/4}\,,
\end{equation}
an inflaton value at horizon exit of order $\phi_*\sim\mathcal{O}(0.2)$ might imply that $\tau_\phi$ needs to be close to $\tau_b$ to realise enough e-foldings to inflation. One might therefore wonder whether this regime is too close to the boundary of the K\"ahler cone where the effective field theory ceases to be valid. However it can be checked in an explicit CY example from \cite{Cicoli:2012vw} that the EFT is always under control. The model is characterised by a volume of the form:
\begin{equation}
\mathcal{V}=\frac19\sqrt{\frac23}\left(\tau_b^{3/2}-\sqrt{3}\,\tau_s^{3/2}-\sqrt{3}\,\tau_\phi^{3/2}\right),
\end{equation}
with:
\begin{equation}
\tau_b=\frac{27}{2}\,t_b^2\,,\qquad\tau_s=\frac92\,t_s^2\,,\qquad\tau_\phi=\frac92\, t_\phi^2\,,    
\end{equation}
and K\"ahler cone conditions:
\begin{equation}
t_b+t_s>0\,,\qquad t_b+t_\phi>0\,,\qquad t_s<0\,,\qquad t_\phi<0\,.
\label{KahlerCone}
\end{equation}
The canonical normalisation looks like:
\begin{equation}
\tau_\phi=\left(\frac{\sqrt{3}}{4}\right)^{2/3}\mathcal{V}^{2/3}\phi^{4/3}\simeq\left(\frac{1}{18\sqrt{2}}\right)^{2/3}\tau_b\,\phi^{4/3}\,,
\end{equation}
implying that at horizon exit:
\begin{equation}
\frac{|t_\phi|}{t_b}=\left(\frac{1}{2\sqrt{6}}\right)^{1/3}\phi_*^{2/3}\simeq 0.6\,\phi_*^{2/3}\simeq 0.2\quad\text{for}\quad\phi_*\simeq 0.2\,.
\end{equation}
As can be seen from the conditions (\ref{KahlerCone}), this region is well inside the K\"ahler cone, signalling that the EFT is valid throughout the whole inflationary dynamics.

\subsection{Number of e-foldings from post-inflation}

The number of e-foldings necessary to solve the horizon problem for the Planck reference pivot scale $k_*=0.05\,\rm{Mpc}^{-1}$ is given by:
\begin{equation}
N_e\simeq 57+\frac14\ln r-\frac14 N_\phi-\frac14 N_\chi+\frac14\ln\left(\frac{\rho_*}{\rho_{\rm end}}\right),    
\label{Ne}
\end{equation}
where $N_\phi$ and $N_\chi$ represent the number of e-foldings of matter domination due to oscillations around the minimum of the canonically normalised inflaton $\phi$ and volume mode $\chi$. Let us stress that the value of $N_e$ given by (\ref{Ne}) is the one relevant to match observations even if the total number of efoldings of inflation can be larger. 

Given that in these scenarios preheating is not efficient enough \cite{Barnaby:2009wr,Antusch:2017flz}, Standard Model (SM) degrees of freedom are produced via perturbative reheating driven by the decay of the longest-lived modulus $\varphi$ which can be either $\varphi\equiv\phi$ or $\varphi\equiv\chi$ depending on the details of the SM realisation with D-branes. Note that $\varphi$ has always a non-zero branching ratio into ultra-light volume axions $\theta_b$ which behave as dark radiation, and so contribute to $\Delta N_{\rm eff}$. This quantity is however strongly constrained by Planck and BAO data: $\Delta N_{\rm eff}\lesssim \mathcal{O}(0.3)$ at $95\%$ CL \cite{Planck:2018vyg}. To get a precise prediction for $\Delta N_{\rm eff}$ one has to compute the width for the decays of $\varphi$ into bulk axions, $\Gamma_{\varphi\to\theta_b\theta_b}$, and into SM particles, $\Gamma_{\varphi \to\text{SM}}$. Given that this last decay width depends on the SM realisation, let us discuss this issue in the next subsection.

\subsection{Standard Model realisation}

In type IIB compactifications, non-Abelian gauge symmetries and chiral matter arise on either magnetised D7-branes or D3-branes at singularities. Note that the SM stack of D7-branes cannot wrap the $\tau_s$ and $\tau_\phi$ divisors. In the case of $\tau_s$, the reason is the tension between non-perturbative effects and chirality \cite{Blumenhagen:2007sm}, while for $\tau_\phi$ is the fact that a $\tau_\phi$-dependent Fayet-Iliopoulos (FI) term would emerge making $\tau_\phi$ too heavy to drive inflation. Thus we need to consider a Calabi-Yau with additional blow-up cycles that can support the SM D7-branes. For concreteness, let us introduce two additional intersecting blow-ups, $\tau_{\rm SM}$ and $\tau_{\rm int}$, with the SM branes wrapped on $\tau_{\rm SM}$ and an internal volume of the form:
\begin{equation}
\mathcal{V}=\tau_b^{3/2}-\tau_s^{3/2}-\tau_\phi^{3/2}-\tau_{\rm SM}^{3/2}-\lambda\left(\tau_{\rm int}-\tau_{\rm SM}\right)^{3/2}\,.
\end{equation}
In the absence of non-zero vacuum expectation values of charged matter fields, D-term stabilisation implies a vanishing FI-term which, in turn, fixes one blow-up modulus in terms of the other as follows: 
\begin{equation}
\xi_{\rm FI}=0\quad\Rightarrow\quad \tau_{\rm SM}=\lambda^2\left(\tau_{\rm int}-\tau_{\rm SM}\right).
\end{equation}
There are then two cases:
\begin{enumerate}
\item If $\lambda=0$, $\tau_{\rm SM}\to 0$, yielding a SM realisation on 
D3-branes at a CY singularity.
\item If $\lambda\neq 0$, $\xi_{\rm FI}=0$ can be seen as fixing $\tau_{\rm int}$ in terms of $\tau_{\rm SM}$, so that $\tau_{\rm SM}$ remains as a flat direction which can be lifted by a loop-generated potential \cite{Cicoli:2011qg}:
\begin{equation}
V(\tau_{\rm SM})=\left(\frac{d_{\rm loop}}{\sqrt{\tau_{\rm SM}}}-\frac{g_{\rm loop}}{\sqrt{\tau_{\rm SM}}-\sqrt{\tau_{\rm s}}}\right)\frac{W_0^2}{\mathcal{V}^3}\,,
\end{equation}
that has a minimum at:
\begin{equation}
\tau_s=\left(1+\sqrt{\frac{g_{\rm loop}}{d_{\rm loop}}}\right)^2\tau_{\rm SM}\sim \tau_{\rm SM}\sim \mathcal{O}(10)\sim g_{\rm SM}^{-2}\,.
\end{equation}
This can give a viable scenario where the SM D7-branes are wrapped around $\tau_{\rm SM}$.
\end{enumerate}

\subsection{Moduli decay rates}

The first step to compute the moduli decay rates is the determination of their mass spectrum. The canonically normalised inflaton $\tau_\phi$ is $\phi$, while $\mathcal{V}$ becomes $\chi$ and their masses read:
\begin{equation}
m_\phi\simeq\frac{W_0\ln\mathcal{V}}{\mathcal{V}}\,M_p   \qquad\text{and}\qquad m_\chi\simeq \frac{W_0}{\mathcal{V}^{3/2}\sqrt{\ln\mathcal{V}}}\,M_p\,.
\end{equation}

\subsubsection*{Volume decay rates}

The dominant decay rates of the volume mode $\chi$ are given by:
\begin{enumerate}
\item Decay into closed string axions $\theta_b$ arising from the kinetic Lagrangian \cite{Cicoli:2012aq}:
\begin{equation}
\Gamma_{\chi\to\theta_b\theta_b}\simeq\frac{1}{48\pi}\frac{m_\chi^3}{M_p^2}\,.
\end{equation}
        
\item Decay into MSSM Higgses $H_u$ and $H_d$ induced by a Giudice-Masiero term in the K\"ahler potential \cite{Cicoli:2012aq}:            
\begin{equation}
\Gamma_{\chi\to H_u H_d}\simeq 2 Z^2\,\Gamma_{\chi\to\theta_b\theta_b}\,,
\end{equation}
where $Z$ is expected to be an $\mathcal{O}(1)$ coefficient. In what follows, we shall set $Z=2$ which helps in some cases to avoid axionic dark radiation overproduction.

\item Decay into SM Higgses $h$ induced by radiative corrections to the Higgs mass \cite{Cicoli:2022fzy}:
\begin{equation}
\Gamma_{\phi\to hh}\simeq\frac{c_{\rm loop}^2}{32\pi}\left(\frac{m_0}{m_\chi}\right)^4\frac{m_\chi^3}{M_p^2}\,,
\end{equation}
where $c_{\rm loop}$ is a loop coefficient and $m_0$ the soft scalar mass. 
\end{enumerate}
The ratio between the volume decay into SM Higgses and the one into bulk axions is given by:
\begin{equation}
\frac{\Gamma_{\chi\to hh}}{\Gamma_{\chi\to\theta_b\theta_b}} \simeq c_{\rm loop}^2\left(\frac{m_0}{m_\chi}\right)^4\,,
\end{equation}
and can take different values depending on the SM realisation:
\begin{enumerate}
\item SM on D7-branes: in this case, supersymmetry breaking gives rise to soft masses of order the gravitino mass, and so:
\begin{equation}
m_0\simeq\frac{W_0}{\mathcal{V}}\gg m_\chi\quad\Rightarrow\quad\frac{\Gamma_{\chi\to hh}}{\Gamma_{\chi\to\theta_b\theta_b}} \simeq \left(c_{\rm loop}\,\mathcal{V}\right)^2\gg 1\,,
\end{equation}
implying that $\chi$ decays mainly into SM Higgses $h$.

\item SM on D3-branes: in this case, the SM is geometrically sequestered from the sources of supersymmetry breaking in the bulk, resulting in suppressed soft terms:
\begin{equation}
m_0\lesssim m_\chi\quad\Rightarrow\quad\frac{\Gamma_{\chi\to hh}}{\Gamma_{\chi\to\theta_b\theta_b}} \lesssim c_{\rm loop}^2\ll 1\,,
\end{equation}
\end{enumerate}
and so $\chi$ decays mainly into $\theta_b$ axions, $H_u$ and $H_d$.

\subsubsection*{Inflaton decay rates}

Let us now turn to the discussion of the inflaton decay rates. We need to consider two cases separately:
\begin{enumerate}
\item Inflaton wrapped by hidden D7-branes: in this case, the main decay channel is into hidden gauge bosons $\gamma_h$:
\begin{equation}
\Gamma_{\phi\to \gamma_h \gamma_h}\simeq\frac{\mathcal{V}}{64\pi} \frac{m_\phi^3}{M_p^2}\,.   
\end{equation}

\item Inflaton not wrapped by any D7-brane: this case features several decay channels:
\begin{itemize}
\item decay into volume moduli $\chi$ and $\theta_b$ axions:
\begin{equation}
\Gamma_{\phi\to\chi\chi}   \simeq\Gamma_{\phi\to\theta_b\theta_b}\simeq \frac{\left(\ln\mathcal{V}\right)^{3/2}}{64\pi\mathcal{V}}\frac{m_\phi^3}{M_p^2}\,.
\end{equation}
If the SM is on D7-branes, $\chi$ then decays instantaneously into $h h$. On the other hand, if the SM is on D3-branes, $\chi$ decays later on into $H_u$ and $H_d$, and $\theta_b\,\theta_b$.

\item When the SM is on D7-branes, one has to take into account extra decays into SM gauge bosons $\gamma$, $\tau_{\rm SM}$ moduli and $\theta_{\rm SM}$ (QCD) axions:          
\begin{eqnarray}
\Gamma_{\phi\to \tau_{\rm SM}\tau_{\rm SM}}  &\simeq& \Gamma_{\phi\to \theta_{\rm SM}\theta_{\rm SM}}\simeq \Gamma_{\phi\to \chi\chi}\,, \\
\Gamma_{\phi\to\gamma\gamma } &\simeq& N_g \Gamma_{\phi\to \chi\chi}\simeq 12\,\Gamma_{\phi\to\chi\chi}\,.
\end{eqnarray}        
In turn, $\tau_{\rm SM}$ then decays instantaneously into $\gamma \gamma$, and $\theta_{\rm SM} \theta_{\rm SM}$ with \cite{Cicoli:2022fzy}:
\begin{equation}
\frac{\Gamma_{\tau_{\rm SM} \to\gamma\gamma}}{\Gamma_{\tau_{\rm SM}\to \theta_{\rm SM} \theta_{\rm SM}}}\simeq 8 N_g \geq 96\gg 1\,.
\end{equation}
\end{itemize}
\end{enumerate}

\subsection{Predictions for cosmological observables}

Using the moduli mass spectrum and decay rates, one can compute the exact number of e-foldings of inflaton and volume domination for each possible brane setup and SM realisation. Plugging their values in (\ref{Ne}) one can then infer the number of e-foldings required to solve the horizon problem, and then substitute $N_e$ in (\ref{ns}) and (\ref{r}) to obtain the predictions for $n_s$ and $r$. Moreover, the knowledge of the moduli decay rates allows to check that the contribution to $\Delta N_{\rm eff}$ from ultra-light bulk axions does not exceeds current observational bounds. The predictions for different model building options are presented in Tab. \ref{tab1}.

\begin{table}[h]
\begin{tabular}{@{}llllllll@{}}
\toprule
Model & $N_e$ & $N_\phi$ & $N_\chi$ & $n_s$ & $r$  & $T_{\rm rh}$ (GeV) & $\Delta N_{\rm eff}$ \\
\midrule
SM on D7s and hidden D7s on $\tau_\phi$ & $53$ & $1$ & $3$ & $0.9765$ & $1.7\times 10^{-5}$  & $4\times 10^{10}$   & $0$  \\
SM on D7s and unwrapped $\tau_\phi$ & $52$ & $8$ & $0$ & $0.9761$   & $1.7\times 10^{-5}$  & $3\times 10^{12}$   & $0.14$  \\
SM on D3s and hidden D7s on $\tau_\phi$ & $51.5$ & $1$ & $10.5$ & $0.9757$   & $1.8\times 10^{-5}$  & $1\times 10^8$   & $0.36$  \\
SM on D3s and unwrapped $\tau_\phi$ & $51.5$ & $11$ & $0.5$ & $0.9757$   & $1.8\times 10^{-5}$  & $1\times 10^8$   & $0.36$  \\
\botrule
\end{tabular}
\caption{Predictions for the main cosmological observables depending on the model-dependent realisation of the Standard Model and inflationary brane setup.}
\label{tab1}
\end{table}

As anticipated, the number of e-foldings of inflation is in the small window $51.5\lesssim N_e \lesssim 53$, resulting in predictions for $n_s$ and $r$ that are not very sensitive to the underlying model building details. What changes more significantly are instead the predictions for the reheating temperature and the amount of extra dark radiation.

\section{Dark Energy}
\label{DarkEnergy}

\subsection{De Sitter from string theory?}

The search for 4D de Sitter (dS) vacua in string compactifications is a crucial problem with both observational and theoretical implications. Even if fully stable dS solutions do not seem to exist, metastable dS vacua could be a viable option. However the difficulty to obtain them within an effective field theory that is under robust theoretical control has raised some doubts about the validity of these solutions. 

According to the extreme point of view of the No dS conjecture \cite{Obied:2018sgi}, metastable dS may be incompatible with quantum gravity, implying that dark energy has to be quintessence. On the other hand, a more conservative viewpoint suggests that the No dS conjecture should be valid only at boundary of moduli space, implying that metastable dS vacua could exist in the interior of moduli space. From this perspective, dS solutions cannot be achieved with parametric control but metastable dS vacua with numerical control could arise thanks to small parameters like $W_0\ll 1$ in KKLT constructions \cite{Giddings:2001yu,Kachru:2003aw} and $\mathcal{V}^{-1}\sim e^{-1/g_s} \ll 1$ in LVS models \cite{Balasubramanian:2005zx,Cicoli:2018kdo}.

It is worth pointing out that AdS solutions are easier to find since dS necessarily breaks supersymmetry. However, several uplifting mechanisms have been proposed in the literature: anti-D3-branes \cite{Kachru:2003aw}, D-terms \cite{Burgess:2003ic}, T-branes \cite{Cicoli:2015ylx}, $\alpha'$ effects \cite{Westphal:2006tn}, non-zero F-terms of the complex structure moduli \cite{Gallego:2017dvd}, non-perturbative effects at singularities \cite{Cicoli:2012fh}. Not all of them have been scrutinized in detail, and it is not so unreasonable to expect that one of them, or a combination of some of them, might be responsible to achieve a metastable dS vacuum. Moreover, recent progress has been made in understanding and classifying quantum corrections to the low-energy effective action at different orders in $\alpha'$ and $g_s$ using higher-dimensional symmetries \cite{Burgess:2020qsc, Cicoli:2021rub}. These studies can help in increasing the computational control over dS solutions. In addition, globally consistent Calabi-Yau models with a SM-like sector on D3-branes and dS from T-branes have been proposed in \cite{Cicoli:2013cha,Cicoli:2021dhg}, providing explicit models where computational issues can be addressed in detail. Finally, it is important to stress that metastable dS vacua can be phenomenologically viable only if their lifetime is longer than the age of the universe. This is a crucial aspect to check for any model that aspires to describe the real world.

\subsection{Quintessence from string theory?}

Contrary to the more conservative case of a cosmological constant, dark energy could be dynamical, as in the prototypical situation of quintessence where a (pseudo)-scalar field rolls down its potential. This cosmological scenario might be singled out if indeed dS vacua will turn out to be in the swampland, or if it will happen to be preferred by observations. 

It is therefore essential to analyse quintessence model building in string cosmology. Due to the difficulty to find saxion quintessence slow-roll down a shallow potential (see however \cite{Cicoli:2012tz}) and axion quintessence with a trans-Planckian decay constant, $f_a\gtrsim M_p$, (as implied also by the weak gravity conjecture \cite{Arkani-Hamed:2006emk}), alternative options could be saxion/axion hilltop quintessence around a dS maximum in a model with a Minkowski/AdS vacuum or a saxion runaway. This last case could involve only tree-level physics, allowing in principle a dark energy solution with parametric control. We shall however argue that only axion hilltop quintessence has any sensible hope to be a viable description of dark energy.

\subsection{No quintessence at the boundary of moduli space}

Let us first consider the possibility to realise quintessence via a saxion runaway at the boundary of moduli space. We shall focus on the volume mode of type IIB string compactifications (similar considerations apply to type IIA and heterotic strings) \cite{Cicoli:2021fsd}. The tree-level K\"ahler potential reads:
\begin{equation}
K=-3\ln\tau\,,
\label{Ktree}
\end{equation}
implying a canonical normalisation of the form (setting $M_p=1$):
\begin{equation}
\mathcal{L}_{\rm kin}=\frac{3}{4\tau^2}\partial_\mu\tau\partial^\mu\tau=\frac12\partial_\mu\phi\partial^\mu\phi    \quad\Leftrightarrow\quad \tau= e^{\sqrt{2/3}\phi}\,.
\end{equation}
Considering the boundary of moduli space where the $\alpha'$ expansion is parametrically under control corresponds to taking the limit $\tau\to\infty$ where the scalar potential can be very well approximated by the tree-level one:
\begin{equation}
V=e^K\left(|D_U W|^2+|D_S W|^2\right) = \frac{V_0}{\tau^3}\,.
\end{equation}
This expression has been derived using the no-scale property of the tree-level K\"ahler potential (\ref{Ktree}) and the fact that the superpotential does not depend on $\tau$ at perturbative level, i.e. $\partial_\tau W=0$. If the complex structure moduli $U$ and the axio-dilaton $S$ are fixed supersymmetrically at leading order, i.e. $D_U W= D_S W=0$, the first non-vanishing contribution to $V$ arises from the leading quantum correction which, for $\tau\to\infty$, can be parametrised as:
\begin{equation}
V = \frac{V_0}{\tau^{3+p}} =V_0\,e^{\lambda\phi} \qquad\lambda=\sqrt{6}(1+p)\qquad p>0\,.
\end{equation}
It is easy to see that this runaway potential does not support an epoch of accelerated expansion since:
\begin{equation}
\epsilon = \frac12\left(\frac{V_\phi}{V}\right)^2 = \frac{\lambda^2}{2} = 3\,(1+p)^2>1\quad\text{for}\quad p>0\,.
\end{equation}
Similar results hold for the case with a dilaton runaway where $g_s\to \infty$, implying that quintessence cannot be realised also in the limit where the string loop expansion is parametrically under control.

\subsection{Multifield quintessence?}

Even if single-field quintessence is not realisable at the boundary of moduli space, multifield quintessence, a more common example in a UV complete setup, could still work, in particular due to the kinetic coupling between a modulus (a saxion) and its associated axion. In fact, it has been shown that a non-geodesic motion in a curved field space can give rise to an epoch of accelerated expansion \cite{Cicoli:2020cfj,Cicoli:2020noz}. The idea is to consider a complex modulus $T=\tau+i\theta$ and the associated kinetic coupling between $\tau$ and $\theta$:
\begin{equation}
\mathcal{L}_{\rm kin}\supset \frac{3}{4\tau^2}\,\partial_\mu\theta\partial^\mu\theta = \frac34\,e^{-2\sqrt{\frac23}\phi}\,\dot{\theta}^2\,.  
\end{equation}
If $\dot{\theta}\neq 0$, this yields an effective time-dependent contribution to the scalar potential of the canonically normalised saxion $\phi$:
\begin{equation}
V_{\rm eff} = V_0\,e^{-\lambda\phi} - \frac34\,e^{-2\sqrt{\frac23}\phi}\,\dot{\theta}^2\,,  
\end{equation}
with the equations of motion that imply $\dot{\theta}^2\sim a^{-6}$ if $m_\theta\simeq 0$. The time-dependent contribution acts effectively as a friction that slows the field $\phi$ down so that $V_{\rm eff}$ can yield an accelerated expansion even if the time-independent potential is very steep. However, concrete string theory examples do not feature accelerating solutions that can match observations. In fact, the supergravity K\"ahler potential of generic string compactifications takes the form:
\begin{equation}
K=-p\ln\left(X+\overline{X}\right),
\end{equation}
with typical values of $p$ being integers like $p=1$ for the dilaton and $p=3$ for the volume mode. For a constant superpotential, the resulting tree-level effective scalar potential looks like: 
\begin{equation}
V_{\rm eff} = V_0\,e^{-\sqrt{2p}\,\phi} - \frac{p}{4} \,e^{-2\sqrt{\frac{2}{p}}\phi}\,\dot{\theta}^2\,.   
\end{equation}
Analysing the fixed points of the resulting dynamical system, it is easy to realise that, starting from a matter dominated universe, the system never features even a transient with an accelerated expansion that matches data with $\omega\simeq -1$ and $\Omega\simeq 0.7$ \cite{Brinkmann:2022oxy}.

\subsection{Challenges for quintessence}
\label{SecChall}

The analysis of the previous sections implies that quintessence, as dS, has to be in the bulk of moduli space where one can hope to have at best numerical, but not parametric, control over the approximations. Hence quintessence model building shares the same challenges as dS constructions, with however extra issues typical just of dynamical dark energy models like \cite{Hebecker:2019csg,Cicoli:2021skd}:
\begin{enumerate}
\item Ultra-light quintessence field: the mass of the quintessence field is typically of the same order as the Hubble scale today, leading to an ultra-light scalar. This can be seen from the $\eta$ parameter:
\begin{equation}
\eta=M_p^2\,\frac{V_{\phi\phi}}{V}\sim \left(\frac{m_\phi}{H_0}\right)^2\lesssim 1\qquad\Rightarrow\qquad m_\phi\lesssim H_0\sim 10^{-60}\,M_p\,.
\end{equation}
The smallness of the mass raises two questions: ($i$) How can such a tiny mass be stable against radiative corrections? ($ii$) How can fifth forces mediated by such an ultra-light scalar be avoided? In fact, present observational bounds require $m_\phi\gtrsim 10^{-4}$ eV for standard moduli which couple to ordinary matter with Planckian strength \cite{Hees:2018fpg}. Note that blow-up moduli, which are geometrically separated in the extra dimensions from the SM, could couple with weaker than Planckian strength for large values of the Calabi-Yau volume \cite{Acharya:2018deu}, even if radiative corrections to $m_\phi$ would still pose a strong challenge.

\item In a string compactification all energy scales are correlated. It is therefore crucial to make sure that the requirement to obtain a tiny Hubble scale today, $H_0\sim \sqrt{V}/M_p\sim 10^{-60}\, M_p$, does not push other scales below current observation bounds. In particular, both the string mass $M_s$ and the scale of the soft supersymmetry breaking terms $M_{\rm soft}$ should lie above the TeV scale: $M_s\gtrsim\mathcal{O}(1-10)$ TeV and $M_{\rm soft}\gtrsim\mathcal{O}(1-10)$ TeV. The first requirement sets a strong upper bound on the value of the volume mode $\mathcal{V}$:
\begin{equation}
M_s\simeq\frac{M_p}{\sqrt{\mathcal{V}}} \gtrsim 10^{-30}\,M_p\qquad\Rightarrow\qquad\mathcal{V}\lesssim 10^{30}\,.
\end{equation}
This observation leads to the following question: Can the observed dark energy scale be reproduced for $\mathcal{V}\lesssim 10^{30}$?

\item The volume mode should be heavy enough to avoid any cosmological moduli problem, requiring $m_{\mathcal{V}}\gtrsim 30$ TeV. Even assuming that the cosmological moduli problem can somehow be avoided (for example via a small initial misalignment of the volume mode or by an enhanced $\mathcal{V}$ coupling to ordinary matter), there should still be no fifth force mediated by the volume mode, which boils down to requiring $m_{\mathcal{V}}\gtrsim 0.1$ meV for standard gravitational couplings to SM particles \cite{Hees:2018fpg}. Note that it is very hard to avoid fifth forces mediated by $\mathcal{V}$ by invoking either screening effects or a geometrical sequestering of the visible sector in the extra dimensions \cite{Acharya:2018deu}.

Hence, generically, a quintessence model free of fifth forces mediated by $\mathcal{V}$ seems to require a large hierarchy between $\mathcal{V}$ and $\phi$: $m_{\mathcal{V}}\gtrsim 10^{-3}\,{\rm eV}\gg m_\phi\sim 10^{-33}\,{\rm eV}$. This can be achieved only if the leading order effects which generate a potential for the volume mode lift only $\mathcal{V}$ but not $\phi$: 
\begin{equation}
V=V_{\rm lead}(\mathcal{V})+V_{\rm sub}(\mathcal{V},\phi)\,,
\end{equation}
with the following hierarchy:
\begin{equation}
\frac{V_{\rm sub}}{V_{\rm lead}}\sim\left(\frac{m_\phi}{m_{\mathcal{V}}}\right)^2 \lesssim 10^{-60}\,.
\label{Hierarchy}
\end{equation}
It is then straightforward to realise that this hierarchy is in strong tension with the lower bound $M_s\gtrsim\mathcal{O}(1-10)$ TeV if $V_{\rm sub}$ is generated by perturbative effects. As a concrete example to illustrate this tension, one can compare $\mathcal{O}(\alpha'^3)$ effects that depend just on $\mathcal{V}$ with string loop corrections which are expected to depend on both $\mathcal{V}$ and $\phi$:
\begin{equation}
\frac{V_{\rm loop}(\mathcal{V},\phi)}{V_{\alpha'^3}(\mathcal{V})} \sim\frac{1}{\mathcal{V}^{1/3}}\lesssim 10^{-60}\quad\Leftrightarrow\quad\mathcal{V}\gtrsim 10^{180}\quad\Rightarrow\quad M_s\ll 1\,\text{TeV}\,.
\end{equation}
\end{enumerate}

\subsection{Quintessence model building}

As we have seen in the previous section, building quintessence models is even more challenging than constructing dS vacua due to the lack of parametric control in both cases and the presence of extra challenges (radiative stability, fifth forces, correct mass scales) for dynamical dark energy \cite{Cicoli:2021skd}. From this analysis, dS vacua appear easier to build but quintessence might happen to be preferred by observations (recent hints might indeed come from DESI \cite{DESI:2025zgx}). 

If that will indeed turn out to be the case, the analysis of Sec. \ref{SecChall} implies that the best candidates to drive dynamical dark energy are axion-like fields since they feature the right properties to overcome the main model building challenges:
\begin{itemize}
\item Being pseudo-scalars that mediate spin-dependent interactions, axions, even if ultra-light, do not mediate any long-range fifth force between macroscopic unpolarised objects.

\item Axions enjoy a continuous shift symmetry that is exact at perturbative level, guaranteeing that their mass is radiatively stable.

\item Axions acquire a mass only via exponentially suppressed non-perturbative effects, allowing to generate a small Hubble scale today without lowering, at the same time, also other relevant scales as the mass of the volume mode, the string scale and the supersymmetry breaking scale.
\end{itemize}
The above considerations suggest that a working quintessence model should feature a leading order potential, $V_{\rm lead}(\mathcal{V})$, that at perturbative level stabilises the volume mode $\mathcal{V}$ while leaving the axion $\phi$ as a flat direction. The axion would then be lifted at subdominant non-perturbative level by $V_{\rm sub}(\mathcal{V},\phi)$. Given that matching observations requires that the axion potential is a tiny correction of order the cosmological constant, $V_{\rm lead}$ should have a supersymmetry breaking Minkowski vacuum. If $V_{\rm lead}$ is generated by $\mathcal{O}(\alpha'^3)$ effects as in LVS models, the hierarchy between $V_{\rm lead}$ and $V_{\rm sub}$ can easily satisfy (\ref{Hierarchy}) for relatively small values of $\mathcal{V}$ that keep $M_s\gg 1\,{\rm TeV}$:
\begin{equation}
\frac{V_{\rm sub}}{V_{\rm lead}}\sim\left(\frac{m_\phi}{m_{\mathcal{V}}}\right)^2 \sim \mathcal{V}^3\,e^{-a\mathcal{V}^{2/3}} \lesssim 10^{-60}\quad\text{for}\quad\mathcal{V}\ll 10^{30}\quad\text{and}\quad M_s\gg 1\,{\rm TeV}\,.
\end{equation}
Despite all these advantages, axions have the big problem that their typical potential:
\begin{equation}
V_{\rm sub}(\mathcal{V},\phi) =\Lambda(\mathcal{V})\left[1-\cos\left(\frac{\phi}{f(\mathcal{V})}\right)\right],
\end{equation}
can drive an epoch of accelerated expansion for arbitrary field values only if their $\mathcal{V}$-dependent decay constant is trans-Planckian, $f(\mathcal{V})\gtrsim M_p$. However, such a large decay constant cannot be obtained in any effective field theory that is under control (and is forbidden by the weak gravity conjecture applied to axions \cite{Arkani-Hamed:2006emk}). Model building way-outs involve multiple axions and alignment mechanisms which however turn out to be rather contrived and hardly under theoretical control. Remaining with a single axion, another possibility is to focus on the more natural case with $f(\mathcal{V})<M_p$ where quintessence can be realised as in hilltop models with the axion around the maximum of its potential.

\subsection{An axion hilltop model}

Let us present a concrete working model of axion hilltop quintessence in string theory exploiting the volume axion of the simplest vanilla LVS model \cite{Cicoli:2024yqh}. The low-energy supergravity theory is characterised by two K\"ahler moduli, $T_b=\tau_b+i\theta_b$ and $T_s=\tau_s+i\theta_s$, and a K\"ahler potential and a superpotential of the form:
\begin{equation}
K=-2\ln\left(\mathcal{V}+\frac{\xi}{g_s^{3/2}}\right)   \qquad\text{and}\qquad W=W_0 + A_s\, e^{-a_s T_s} + A_b\,e^{-a_b T_b}\,, 
\end{equation}
where $\xi$, $W_0$, $A_s$, $A_b$, $a_s$ and $a_b$ are all $\mathcal{O}(1)$ constants, $g_s\lesssim 0.1$ and the Calabi-Yau volume $\mathcal{V}$ looks like:
\begin{equation}
\mathcal{V}=\tau_b^{3/2}-\tau_s^{3/2}\simeq \tau_b^{3/2}\qquad\text{for}\qquad \tau_b\gg \tau_s\,.
\end{equation}
At leading order in a large volume expansion, the potential takes the same form as (\ref{Vlead}) and depends on $\tau_b$, $\tau_s$ and $\theta_s$ with a flat $\theta_b$ axion. If the uplifting contribution can be tuned appropriately, it features a Minkowski vacuum with broken supersymmetry at:
\begin{equation}
\langle\theta_s\rangle=0\qquad \langle\tau_s\rangle\sim g_s^{-1}\gtrsim 10\qquad \langle\mathcal{V}\rangle\sim e^{a_s\langle\tau_s\rangle} \sim e^{a_s/g_s}\gg 1\,.
\end{equation}
Subdominant $T_b$-dependent non-perturbative corrections to $W$ lift the axionic direction $\theta_b$ generating a potential for the canonically normalised axion $\phi$ of the form:
\begin{equation}
V_{\rm sub} (\mathcal{V},\phi) \simeq e^{-\sqrt{\frac23}\frac{M_p}{f}}\,M_p^4\left[1-\cos\left(\frac{\phi}{f}\right)\right],
\label{VSub}
\end{equation}
where the axion decay constant is:
\begin{equation}
f=\sqrt{\frac32}\frac{M_p}{a_b\tau_b}\,.
\end{equation}
This potential can drive quintessence if the overall scale of $V_{\rm sub}$ is of order $10^{-120}\,M_p^4$, corresponding to $f\sim 0.0045\,M_p$. Interestingly, this scale of the axion decay constant can be achieved for a natural value of the extra-dimensional volume, $\mathcal{V}\simeq \tau_b^{3/2}\sim \mathcal{O}(10^3)$ for $a_b\sim \mathcal{O}(1)$, which is also large enough to keep computational control over the low-energy effective field theory. Moreover, the mass of the volume mode is well above any phenomenological and cosmological bound since $m_{\mathcal{V}}\sim 10^{13}$ GeV. 

\subsection{Hilltop and initial conditions}

\begin{figure}[h]
\centering
\includegraphics[width=0.75\textwidth]{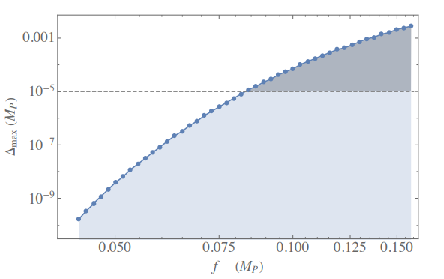}
\caption{Upper limit on distance from the maximum as a function of the decay constant $f$ in order to obtain a phenomenologically viable model of axion hilltop quintessence (plot taken from \cite{Cicoli:2024yqh}).}
\label{NewFig}
\end{figure}

Besides reproducing the correct dark energy scale, a working quintessence model should also match other observations, like $\omega_\phi\simeq -1$ and $\Omega_\phi\simeq 0.7$. The requirement to reproduce these features indicates how close $\phi$ should be to the maximum in axion hilltop quintessence. This is shown in Fig. \ref{NewFig}, taken from \cite{Cicoli:2024yqh}, which implies that sub-Planckian decay constants require tuned initial conditions. This is clearly challenging in the absence of a mechanism that can determine these particular initial conditions dynamically. A model building possibility could exploit the dependence on open string moduli of the prefactor of non-perturbative effects. If the open strings evolve with time, this could result in a time-dependent contribution to the potential of the dark energy axion. In that case, the hilltop region might initially correspond to a minimum that turns then into a maximum due to a non-trivial dynamics of the open string modes. 

Another challenge of these constructions is the fact that, in the absence of time-dependent contributions to the quintessence potential, the dark energy axion is effectively massless during inflation, and so quantum diffusion during this epoch would move $\phi$ by an amount of the order the Hubble scale during inflation, $\Delta\phi\sim H_{\rm inf}$. To preserve good initial conditions determined by Fig. \ref{NewFig} as a function of the axion decay constant $f$, one has therefore to require $H_{\rm inf}\lesssim \Delta_{\rm max}(f)$. This leads to two regimes:
\begin{enumerate}
\item $f\simeq 0.1\,M_p$: in this case the bound becomes:
\begin{equation}
H_{\rm inf}\lesssim 10^{-4}\,M_p\sim 10^{14}\,\text{GeV}\,.    
\end{equation}
This upper limit on the inflationary scale $H_{\rm inf}$ is similar to the one derived from the non-observation of primordial tensor modes, and so does not represent a real obstacle to quintessence model building. However we have seen in the previous section that a dark energy scale of order $10^{-120}\,M_p^4$ would require a much smaller decay constant. In fact, plugging $f\simeq 0.1\,M_p$ in (\ref{VSub}) would lead to $V_{\rm sub}\sim 10^{-5}\, M_p^4$.

\item $f\simeq 0.02\,M_p$: in this case the upper bound on the inflationary scale turns out to be very strong:
\begin{equation}
H_{\rm inf}\lesssim 10^{-18}\,M_p\sim 1\,\text{GeV}\,. 
\label{StrongUpperBound}
\end{equation}
Clearly, it is very challenging to realise inflation at such a low scale. Moreover, it is worth stressing that this bound becomes even stronger for the value of the decay constant, $f\simeq 0.0045$, required to get the observed cosmological constant scale.
\end{enumerate}
Note that the two cases have opposite problems: the first case can easily accommodate a large inflationary scale, $H_{\rm inf}$, but cannot match the current value of the Hubble scale, $H_0$, while the second case can reproduce $H_0$ but cannot allow for a large enough $H_{\rm inf}$. Different model building mechanisms can be implemented to solve these tensions. In the first case, one can exploit the large suppression of poly-instanton effects to generate an axion potential with the observed dark energy scale even for $f\simeq 0.1\,M_p$, whereas in the second case axion alignment can help to relax the strong upper bound (\ref{StrongUpperBound}) thanks to an effective decay constant around $f_{\rm eff}\simeq 0.1\,M_p$.

\subsection{Quintessence from poly-instantons}

Let us present the main ingredients of a working axion quintessence hilltop model \cite{Cicoli:2024yqh} based on poly-instanton effects \cite{Blumenhagen:2008ji,Blumenhagen:2012kz}. The starting point is to consider LVS models with fibred Calabi-Yau threefolds with volume of the form:
\begin{equation}
\mathcal{V}=\sqrt{\tau_f}\tau_b-\tau_s^{3/2}\simeq \sqrt{\tau_f}\tau_b\,.
\end{equation}
The model features an extra K\"ahler modulus, $T_f=\tau_f+i\theta_f$, on top of the two typical K\"ahler moduli, $T_b$ and $T_s$, of vanilla LVS constructions. The scalar potential receives several contributions at different orders in a large volume expansion, and it can be schematically written as:
\begin{equation}
V=V_{\rm lead}(\mathcal{V},\tau_s,\theta_s)+V_{\rm inf}(\tau_f)+V_{\rm sub}(\theta_b,\theta_f)\,,
\end{equation}
where $V_{\rm lead}(\mathcal{V},\tau_s,\theta_s)$ is the leading order LVS potential (\ref{Vlead}) which stabilises $\tau_s$, $\theta_s$ and the combination of $\tau_f$ and $\tau_b$ corresponding to $\mathcal{V}$. On the other hand, $V_{\rm inf}(\tau_f)$ is a loop-generated potential that lifts at subleading order the remaining flat direction in the ($\tau_f,\tau_b$)-plane which we parametrise as $\tau_f$ without loss of generality. Expressing $V_{\rm inf}$ in terms of the canonically normalised modulus $\phi$, this turns out to be the potential of Fibre Inflation that in the inflationary plateau can be approximated as \cite{Cicoli:2008gp}:
\begin{equation}
V_{\rm inf}\simeq V_0\left(1-\frac43\,e^{-\phi/\sqrt{3}}\right)\qquad\text{with}\qquad H_{\rm inf}\simeq 10^{-5}\,M_p\,.
\end{equation}
The model features two ultra-light bulk axions, $\theta_f$ and $\theta_b$, whose potential $V_{\rm sub}(\theta_b,\theta_f)$ is generated by tiny poly-instanton corrections to the superpotential of the form:
\begin{equation}
W= W_0 + A_s\,e^{-a_s T_s} + A_b\,e^{-a_b T_b + e^{-a_f T_f}}\,.  
\end{equation}
The total potential for the two canonically normalised axions $\phi_b$ and $\phi_f$ turns out to scale as:
\begin{equation}
V_{\rm sub}\simeq e^{-a_b\tau_b}\left[1-\cos\left(\frac{\phi_b}{f_b}\right)\right]+e^{-a_b\tau_b-a_f\tau_f}\left[1-\cos\left(\frac{\phi_b}{f_b}+\frac{\phi_f}{f_f}\right)\right].
\end{equation}
The heavier axion $\phi_b$ is a spectator field which can at most contribute to $0.2\%$ of dark matter, while $\phi_f$ drives quintessence. After fixing $\phi_b=0$, the quintessence potential becomes:
\begin{equation}
V_{\rm DE}\simeq e^{-f_f^{-1}-f_b^{-1}}\left[1-\cos\left(\frac{\phi_f}{f_f}\right)\right].   
\end{equation}
As studied in \cite{Cicoli:2024yqh}, in order to have a working model with the right cosmological constant scale and a large enough decay constant $f_f$ to protect the initial conditions against quantum diffusion during inflation, the two decay constants need to take the values:
\begin{equation}
f_f=\frac{N_f}{2\sqrt{2}\pi\tau_f}\,M_p\simeq 0.1\,M_p\qquad\text{and}\qquad f_b=\frac{N_b}{2\pi\tau_b}\,M_p\simeq 0.005\,M_p\,.
\end{equation}
These values can be easily achieved choosing:
\begin{equation}
\tau_f\sim\mathcal{O}(5)\,, \qquad\tau_b\sim\mathcal{O}(500)\,,\qquad N_f \sim\mathcal{O}(5)\,,\qquad N_b\sim\mathcal{O}(10)\,.  
\end{equation}
In turn, the masses of the two axions turn out to be:
\begin{equation}
m_{\theta_b}\sim 10^{-29}\,\text{eV}\qquad\text{and}\qquad m_{\theta_f}\sim 10^{-32}\,\text{eV}\,.
\end{equation}

\section{Conclusions}
\label{sec13}

In this manuscript we reviewed an entire class of string models of both early and late time cosmology. 

We focused on the K\"ahler sector of no-scale type IIB supergravity, showing that the potential of moduli orthogonal to the volume mode features a plateau region that can drive inflation. The flatness of the inflaton potential is due to an approximate rescaling shift symmetry, while its actual shape depends on the nature of breaking effects (either perturbative or non-perturbative) and the topology of the divisor whose volume is parametrised by the inflaton (either bulk or local cycles).

In these scenarios reheating can naturally occur via the perturbative decay of the lightest modulus which, depending on model building details, can be either the inflaton itself or the volume modulus. This leads to a non-standard cosmological history with early epochs of matter domination where most of the energy density of the universe is stored in moduli oscillating around the minimum of their potential. When the lightest modulus decays, it can produce both Standard Model and hidden sector degrees of freedom, like ultra-light axions, which contribute to dark radiation. 

As an illustrative example, we described more in detail the model of Loop Blow-up Inflation \cite{Bansal:2024uzr}. In this framework, inflation is driven by a blow-up mode whose potential is generated by 1-loop corrections to the K\"ahler potential. The moduli-dependence of these perturbative corrections has been conjectured via an educated extrapolation from toroidal computations and it has been confirmed by effective field theory arguments based on the 1-loop renormalisation of the moduli 2-point function and the Coleman-Weinberg potential. The inflaton potential can be approximated as $
V(\phi)\simeq V_0\left(1-c\,\mathcal{V}^{-1/3}\phi^{-2/3}\right)$ and inflation can take place inside the K\"ahler cone where the effective theory is under control. In order to match observations, the internal volume and the inflaton at horizon exit have to be, respectively, of order $\mathcal{V}\sim \mathcal{O}(10^4)$ and $\phi_*\sim \mathcal{O}(0.2)$. The scalar spectral index is in perfect agreement with CMB data ($0.9757\lesssim n_s\lesssim 0.9765$) while the tensor-to-scalar ratio is predicted to be of order $r\simeq 2\times 10^{-5}$. The post-inflationary evolution is characterised by epochs of moduli domination and reheating from moduli decay. Depending on the Standard Model realisation with either D7- or D3-branes, the number of e-foldings is in the range $51.5\lesssim N_e\lesssim 53$, while the axionic contribution to dark radiation is $0\lesssim \Delta N_{\rm eff} \lesssim 0.36$.

In the second part of this review, we briefly discussed dark energy model building in string theory comparing de Sitter vacua with quintessence models. We pointed out that, similarly to de Sitter solutions, dynamical dark energy models are hard to control from the theoretical point of view and, in addition, face a few phenomenological challenges like fifth-forces, radiative stability and the requirement to get the right cosmological constant scale without lowering other scales below observational bounds.

In fact, no phenomenologically viable quintessence solution seems to appear at the boundary of moduli space where both the $\alpha'$ and the $g_s$ expansions are under control. Kinetically coupled saxion-axion string models can support late time acceleration but without reproducing $\omega_\phi\simeq -1$ and $\Omega_\phi\simeq 0.7$. 

This analysis shows that quintessence models do not seem to provide realistic alternatives to de Sitter vacua which arise more generically in the string landscape. If however dynamical dark energy will turn out to be preferred by observations, we argued that the most promising candidates to drive quintessence are axion-like fields. The reason is that these fields, being pseudo-scalars, naturally evade fifth-force bounds and, enjoying an exact perturbative shift symmetry, do not face any problem with radiative stability of their mass. Moreover, given that the axion potential is exponentially suppressed by non-perturbative effects, it is relatively easy to reproduce a small cosmological constant while keeping the string and the moduli masses above the TeV scale. 

The main problem with axions is however that the simplest axion potential is not steep enough to yield acceleration for an arbitrary axion field value. The way-out seems to be axion hilltop quintessence where the axion sits around the maximum of its potential. Even allowing for a suitable choice of initial conditions, being effectively a flat direction during inflation, quantum diffusion would move the axion away from the maximum, destroying the possibility for the axion to drive a late-time epoch of accelerated expansion. The requirement to avoid this problem sets an upper bound on the Hubble scale during inflation, $H_{\rm inf}$, which depends on the axion decay constant $f$. If $f\simeq 0.1\,M_p$ the bound is $H_{\rm inf}\lesssim 10^{14}$ GeV, while if $f\simeq 0.02\,M_p$ one gets $H_{\rm inf}\lesssim 1$ GeV. When $f\simeq 0.1\,M_p$, the upper bound on $H_{\rm inf}$ is not problematic but the axion decay constant is not small enough to suppress the axion potential down to the observed value of the cosmological constant. We therefore presented a model with two axions where the right dark energy scale can be reproduced exploiting poly-instanton corrections to the superpotential. The model is not tuned but requires an explicit Calabi-Yau example. When instead $f\simeq 0.02\,M_p$, the strong upper bound on $H_{\rm inf}$ could be relaxed by considering a two-axion model with alignment to get a larger effective decay constant of order $f_{\rm eff}\simeq 0.1\,M_p$. Such a model would however look contrived and tuned.

\backmatter

\bmhead{Acknowledgements}

MC would like to express his deep gratitude to all coauthors of the papers reviewed in this manuscript.

\bibliography{sn-bibliography}

\end{document}